\documentclass[prl,twocolumn,amsmath,amssymb]{revtex4}

\usepackage[dvips]{epsfig}
\usepackage[dvips]{graphicx}
\usepackage{dcolumn}
\usepackage{bm}

\graphicspath{{converted_graphics/}}
\begin{document}

\title{Controllable valley splitting in silicon quantum devices}

\author{Srijit Goswami$^{1\dagger}$}
\author{K. A. Slinker$^{1\dagger}$}
\author{Mark Friesen$^{1\dagger}$}
\author{L. M. McGuire$^1$}
\author{J. L. Truitt$^1$} 
\author{Charles Tahan$^2$} 
\author{L. J. Klein$^1$} 
\author{J. O. Chu$^3$}
\author{P. M. Mooney$^4$}
\author{D. W. van der Weide$^5$}
\author{Robert Joynt$^1$} 
\author{S. N. Coppersmith$^{1}$} 
\author{Mark A. Eriksson$^{1*}$} 
\affiliation{$^1$Department of Physics, University of
Wisconsin-Madison, Wisconsin 53706, USA} 
\affiliation{$^2$Cavendish Laboratory, JJ Thomson Ave, Cambridge CB3 0HE, United Kingdom}
\affiliation{$^3$IBM Research Division, T. J. Watson
Research Center, New York 10598, USA}
\affiliation{$^4$Department of Physics, Simon Fraser University, Burnaby, 
British Columbia V5A 1S6, Canada}
\affiliation{$^5$Department of Electrical and Computer Engineering, University of
Wisconsin-Madison, Wisconsin 53706, USA}
\affiliation{$^\dagger$These three authors contributed equally to this work,}
\affiliation{$^*$e-mail: maeriksson@wisc.edu}

\begin{abstract}
Silicon has many attractive properties for quantum computing, and the quantum dot architecture is appealing because of its controllability and scalability.  However, the multiple valleys in the silicon conduction band are potentially a serious source of decoherence for spin-based quantum dot qubits.  Only when these valleys are split by a large energy does one obtain well-defined and long-lived spin states appropriate for quantum computing.  Here we show that the small valley splittings observed in previous experiments on Si/SiGe heterostructures result from atomic steps at the quantum well interface.  Lateral confinement in a quantum point contact limits the electron wavefunctions to several steps, and enhances the valley splitting substantially, up to 1.5~meV.  The combination of electronic and magnetic confinement produces a valley splitting larger than the spin splitting, which is controllable over a wide range.  These results improve the outlook for realizing spin qubits with long coherence times in silicon-based devices.
\end{abstract}


\maketitle

The fundamental unit of quantum information is the qubit.  Qubits can be constructed from the quantum states of physical objects like atomic ions \cite{cirac95}, 
quantum dots \cite{loss98,ciorga00,fujisawa02,elzerman04,johnson05,petta05} 
or superconducting Josephson junctions \cite{shnirman97}.  
A key requirement is that these quantum states should be well-defined and isolated 
from their environment.  An assemblage of many qubits into a 
register and the construction of a universal set of operations, including 
initialization, measurement, and single and multi-qubit gates, would 
enable a quantum computer to execute algorithms for certain difficult computational problems like 
prime factorization and database search far faster 
than any conventional computer \cite{nielsenbook}.

The solid state affords special benefits and challenges for qubit operation and quantum computation.
State-of-the-art fabrication techniques enable the positioning of electrostatic gates with a resolution of several nanometers, paving the way for large scale implementations.  On
the other hand, the solid state environment provides numerous pathways for decoherence
to degrade the computation \cite{cerletti05}.  Spins in silicon offer a special
resilience against decoherence because of two desirable materials properties 
\cite{kane98,yablonovitch03}:  a small spin-orbit coupling and predominately spin-zero nuclei.  
Isotopic purification could essentially eliminate all nuclear decoherence mechanisms.

Silicon, however, also has a property that potentially can increase decoherence.  Silicon has multiple conduction band minima or valleys at the same energy.  Unless this degeneracy is lifted, coherence and qubit operation will be threatened.  In strained silicon quantum wells there are two such degenerate valleys \cite{schaeffler97} whose quantum numbers and energy scales compete directly with the spin degrees of freedom.  In principle, sharp confinement potentials, like the quantum well interfaces, couple these two valleys and lift the degeneracy, providing a unique ground state if the coupling is strong enough \cite{ando82,boykin04}. 
Theoretical analyses for noninteracting electrons in perfectly flat (100) quantum wells
predict a valley splitting of order 1~meV or 10~K \cite{boykin04}.
However, existing data for Si/SiGe quantum wells, obtained so far at high magnetic fields, 
show a small valley splitting.  Extrapolation to low fields suggests a valley splitting of only $\mu$eV, much too small for spintronics applications
\cite{weitz96,koester97,khrapai03,lai04,pudalov}.  

Here we show that valley splitting can be controlled and greatly enhanced by confinement in nanostructures.  Theoretically, we show that atomic steps at a quantum well interface
suppress the valley splitting compared with a flat interface.  This suggests lateral confinement would increase the valley splitting by reducing the number of steps seen by the wavefunction.  We demonstrate experimentally that electronic confinement in nanostructures leads to very large valley splittings that approach the theoretical predictions for flat quantum wells with no steps \cite{boykin04}.  
At all magnetic fields, the valley splitting is much larger than the spin splitting, as required for quantum computing. 
We also probe the effects of magnetic confinement in the absence of electronic confinement by using a wide Hall bar geometry.  A low-field microwave spectroscopy, analogous to electrically detected electron spin resonance \cite{dobers88}, enables us to measure valley splitting in smaller magnetic fields than previously possible.  In the absence of strong lateral confinement, the valley splitting is less than the spin splitting and exhibits a strikingly linear dependence on magnetic field.

\begin{figure}[t] 
  \centering
  \includegraphics[bb=25 345 318 738,width=3in,keepaspectratio]{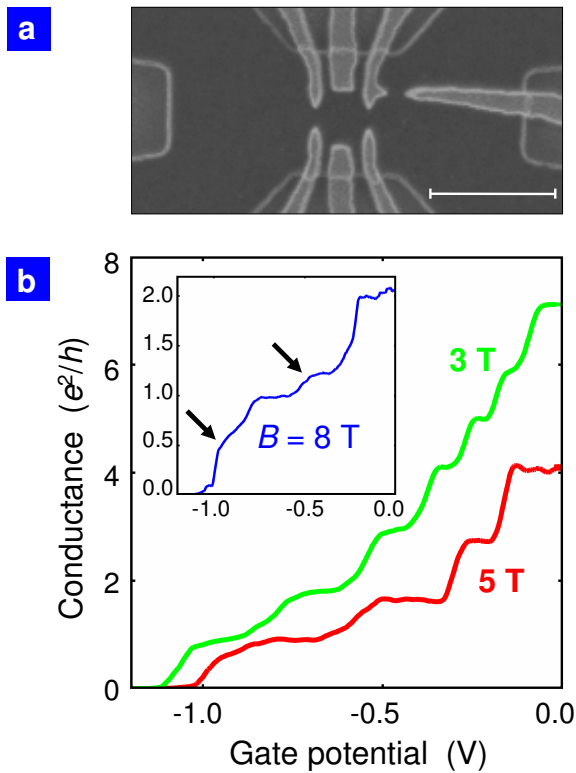}
  \caption{
{\bf Quantum point contact.  a},  The experimental device used in the conduction 
measurements.  Scale bar (lower right) defines 1~$\mu$m.
{\bf b}, the conductance data as a function of gate voltage.  Steps occur every $e^2/h$, 
indicating complete lifting of all degeneracies.  At the magnetic field $B=8 \text{T}$,
the inset shows features reminiscent of the so-called 0.7~structure 
\cite{cronenwett02}}
  \label{fig:VSNPfig1}
\end{figure}

The key physical feature of our samples which is affected by confinement is the 
presence of steps in the quantum well.  The silicon/silicon-germanium quantum wells used in all of our experiments 
were grown on a $2^\circ$ tilt from (100), as is common in commercial wafers.
(Further details on the sample growth are provided in the Methods section.)  
In addition to global tilt, strained 
heterostructures like those used here also exhibit natural roughness and local tilting.  
Ando has suggested that valley splitting 
is strongly suppressed at quantum well interfaces that are tilted with respect to their 
crystallographic axes \cite{ando79}, and we have developed an effective mass theory to calculate the valley splitting
in miscut quantum wells \cite{friesen06}.
Our theory results in the following expression for the valley splitting:
\begin{equation}
E_v = 2\left|\int e^{-2ik_0 z} |F({\bm r})|^2 V_v({\bm r})d^3r \right| . \label{eq:Ev}
\end{equation}
Here, $F({\bm r})$ is a conventional envelope function oriented with respect to the tilted
quantum well.  The phase factor $e^{-2ik_0 z}$ arises from the Kohn-Luttinger effective 
mass approximation \cite{kohn}, and is oriented with respect to the crystallographic 
axis, not the quantum well.  Eq.~(\ref{eq:Ev}) can be understood as an overlap of the
wavefunctions in $k$-space, which are centered on the valley minima $\pm k_0\hat{z}$.
The valley coupling potential $V_v({\bm r})$ is 
nonzero only within several Angstroms of the quantum well interface.  We can 
approximate it as a Dirac $\delta$-function:
\begin{equation}
V_v({\bm r}) =v_0 \delta (z-z_i(x)) , \label{eq:Vv}
\end{equation}
where $v_0$ is the strength of the valley coupling, and the interface position $z_i(x)$
describes the tilted plane of the quantum well.  
For quantum wells with no tilt, Eqs.~(\ref{eq:Ev})
and (\ref{eq:Vv}) agree very well with sophisticated tight 
binding theories \cite{boykin04}.  The theories predict oscillations of the valley splitting 
as a function of quantum well width, due to different phase factors on the top 
and bottom interfaces.  Such interference effects are a recurring theme in valley coupling
\cite{koiller02}, leading to the suppression of $E_v$.

\begin{figure}[t] 
  \centering
  \includegraphics[bb=40 420 374 760,width=2.7in,height=2.7in,keepaspectratio]{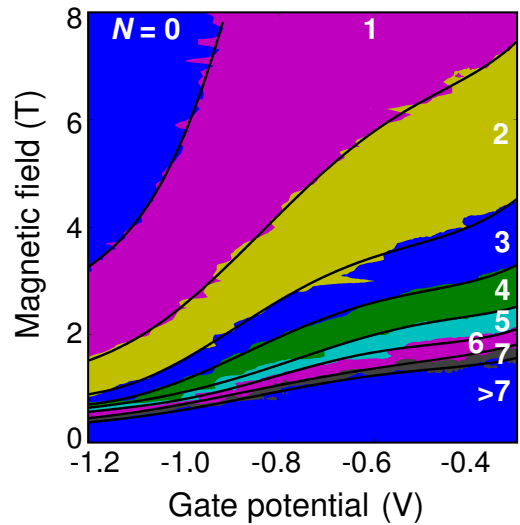}
  \caption{
{\bf Step transitions.}  The step transitions between conductance plateaus in 
Fig.~1b
are mapped out as a function of magnetic field $B$ and gate voltage $V_g$.  The dark lines 
show the results of fitting to Eq.~4, as described in the Supplementary Material.  }
  \label{fig:VSNPfig2}
\end{figure}

Similar interference effects also lead to a strong suppression of the valley splitting 
in a tilted 
quantum well.  To evaluate this effect, we treat the tilted interface 
as a series of atomic steps.  In Eq.~(\ref{eq:Ev}), the phase angles associated
with consecutive steps differ by $\sim 0.85 \pi$, and are nearly out of phase.  
For an electron spread over many steps, the valley splitting is strongly suppressed, 
and vanishes completely in the limit of full 
delocalization.  However, under lateral confinement (\textit{e.g.}, magnetic confinement), 
the electron covers only a finite number 
of steps, leading to an increased valley splitting compared with delocalized electrons.  
In the presence of an 
external confinement potential (\textit{e.g.}, a quantum dot), the magnetic field
and the external potential will both enhance valley splitting, with an approximate form 
given by
\begin{equation}
E_v\simeq \sqrt{\Delta^2_\text{ext}+(\Delta_B B)^2},  \label{eq:Delta}
\end{equation}
where $\Delta_\text{ext}$ is the
valley splitting due to the external potential.

\begin{figure}[t] 
  \centering
  \includegraphics[bb=42 377 565 752,width=3.1in,keepaspectratio]{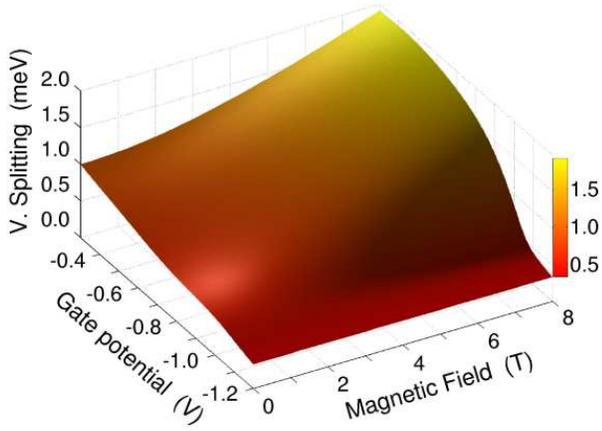}
  \caption{
{\bf Valley splitting.}  A perspective plot of the valley 
splitting in the lowest quantum subband as a function of $B$ and $V_g$.  The splitting is 
large and can be controlled as a function of $B$ and $V_g$. }
  \label{fig:VSNPfig3lr}
\end{figure}

Here we describe our experiments in nanostructured devices in a two-dimensional 
electron gas, demonstrating control of 
valley splitting by confinement.  We first make use of the device shown in Fig.~1a.  
By applying a negative bias to pairs of gates, we deplete the underlying two-dimensional 
electron gas to form quantum point contacts \cite{vanwees88}.  Making use of the fact that
the conductance in 
quantum point contacts is quantized in units of $e^2/h$ times the number of occupied 
subbands, previous analyses have shown how to fit conductance data to a model 
Hamiltonian and extract the {\em orbital} subband energy spacings 
\cite{vanwees91}.

\begin{figure}[t] 
  \centering
  \includegraphics[bb=42 414 364 737,width=2.7in,height=2.7in,keepaspectratio]{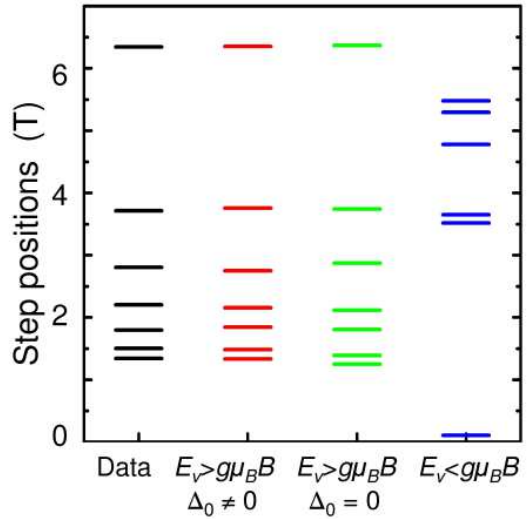}
  \caption{
{\bf Comparison of valley and spin excitations.}  
The experimental step transitions for $V_g=0.5$~V in column 1 are compared with 
fits to Eq.~(4) (columns 2-4) under differing constraints.  Column 2 exhibits good  
agreement with the data when the valley splitting is constrained to be larger than the 
spin splitting.  In contrast, column 4 shows poor agreement under the opposite constraint.  
Column 3 shows an alternative fit, with valley splitting larger than spin splitting, but 
vanishing at zero magnetic field.}
  \label{fig:VSNPfig4lr}
\end{figure}

In this work we extend this approach and show that the valley splitting itself can be extracted from point contact conductance characteristics.  Fig.~1b shows the quantized conductance through a point contact in the device shown.
For nearly degenerate spin and valley states, one would expect conductance steps of four times the conductance quantum ($4e^2/h$).  
However, the steps clearly occur every $e^2/h$, indicating both non-zero spin
and non-zero valley splittings.
We analyze the point contact spectroscopy data as follows.  Building upon 
Eq.~(\ref{eq:Delta}), we write the subband energies
of the quantum point contact as \cite{vanwees91}
\begin{equation}
E_\text{QPC} = (n+1)\sqrt{(\hbar \omega_0)^2+(eB\hbar/2m^*)^2}
+n_B g\mu_BB +n_v E_v+eV_b  \label{eq:EQPC}
\end{equation}
where $E_v$ is given in Eq.~(\ref{eq:Delta}).
The first term in Eq.~(\ref{eq:EQPC}) is the kinetic energy, where $n$ denotes the 
subband index.  The second term 
is the Zeeman spin splitting, and the fourth is the electrostatic potential.  The third term is the valley splitting that we seek.  The indices $n_B,n_v=\pm 1/2$ correspond
to the Zeeman and valley states, respectively.  The electrostatic potential $V_b$ depends 
on the gate voltage $V_g$ but not on $n$.  Since the size of the subband wavefunction 
determines the number of atomic steps that are covered, $\Delta_\text{ext}$ and $\Delta_B$ 
depend on both
$V_g$ and the subband index $n$.  From Eq.~(\ref{eq:EQPC}), we see that $E_\text{QPC}$ varies smoothly 
as a function of the gate voltage and the magnetic field.
However, as $V_g$ and $B$ are varied, the number of modes below the 
Fermi level changes discretely, causing steps in conductance, as observed in Fig.~1b.  

We have measured the quantum point contact conductance as a function of gate voltage and magnetic field (80 magnetic field traces, each at 700 different gate voltages), as shown in Fig.~2.  To fit the transitions between conductance plateaus, the experimental data 
have been mapped to the nearest integer multiples of $e^2/h$, corresponding to the numerical
labels shown \cite{vanwees91}, and the transitions have been smoothed using higher
order polynomials (see figure).
As described in detail in the Supplementary Materials, we fit to Eq.~(\ref{eq:EQPC}) 
and obtain $\Delta_\text{ext}$ and $\Delta_B$, for $n=1$ and $n=2$, and $V_b$, all as a 
function of $V_g$.    The
confinement energy $\hbar \omega_0$ was found to remain essentially constant and very small
for all gate voltages.  To help stabilize the fitting procedure, we fix $\omega_0$ 
to zero.

The resulting valley splitting for the subband with $n=1$ is plotted in Fig.~3,
while the valley splitting for the subband with $n=2$ is shown in the Supplementary 
Materials.  (Data from a second device on the same wafer, not shown,
yield results that are very similar to those shown here.)
The results show that the valley splitting is a tunable function of both electrostatic 
and magnetic confinement.  A crucial question is whether the valley splitting is larger
than the spin splitting.  We have performed many variations of the analysis presented
here.  For example, 
Eq.~(\ref{eq:EQPC}) can be implemented with the valley splitting either larger or smaller 
than the spin splitting.  Fig.~4 shows the best fits obtained using valley splitting 
both larger and smaller than the spin splitting.  The best fit with the valley splitting constrained to be smaller than the spin splitting is shown in the right hand column, and 
it is very poor.  Thus we conclude that the valley splitting is larger than the 
spin splitting in our quantum point contacts.  A more subtle question involves the precise form of Eq.~(\ref{eq:EQPC}).  For example, could the valley splitting be zero at at zero applied magnetic field?  As shown in Fig.~4, the fit with a non-zero intercept is better than the fit with a zero intercept, but not dramatically so.  Thus, we are confident 
that the valley splitting is larger than the spin splitting for even arbitrarily small 
magnetic fields, but we cannot say with complete confidence that the valley splitting is non-zero at zero magnetic field.

\begin{figure}[t] 
  \centering
  \includegraphics[bb=31 225 526 720,width=2.9in,keepaspectratio]{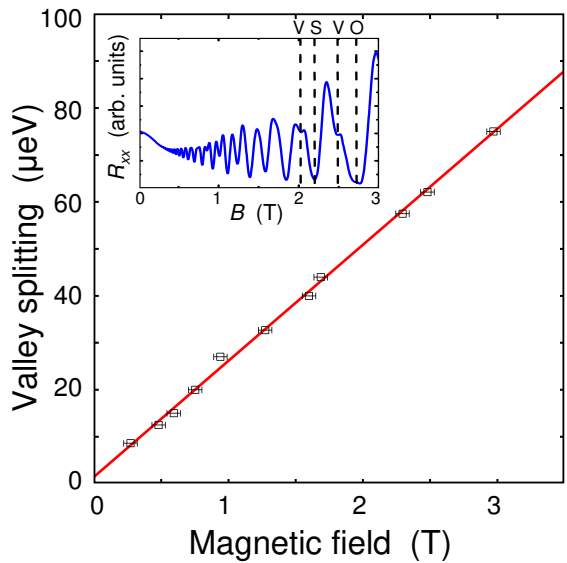}
  \caption{
{\bf Microwave spectroscopy of the valley splitting in a Si/SiGe Hall bar.}  The 
inset shows Shubnikov de Haas oscillations with the valley (V), spin (S), and orbital (O) 
minima labeled.  Magnetic confinement causes the valley splitting to increase as the 
magnetic field increases.  The error bars reflect small shifts in the peak positions with
varying microwave power.}
  \label{fig:VSNPfig5}
\end{figure}

The valley splitting exhibits an interesting dependence on gate voltage as well as on 
magnetic field, decreasing as the device moves closer to pinch-off. Near pinch-off, 
the electron density is low, so many-body enhancements
to the valley splitting are negligible.  We therefore compare the experimentally measured
valley splitting at
pinch-off ($\sim 0.35$~meV) to the predictions of single electron theory. As described
in the Supplementary Information, the resulting prediction for our quantum point contact is 
$E_v=0.46$~meV; because the magnetic
confinement is weaker than the electrostatic confinement, the valley splitting is independent
of magnetic field. 
The fact that the experimental measurement is slightly smaller than the
theoretical value suggests that the few-electron limit is appropriate, but that the presence
of several interfacial steps may suppress the valley splitting slightly compared to a flat well.
In other regions of Fig. 3, where the electron density is larger, the valley splitting is
larger than the theoretical prediction for noninteracting electrons and it exhibits a significant magnetic field dependence.  We ascribe this behavior
to many-body effects, which have been considered elsewhere \cite{ohkawa77}, 
but are not yet
fully understood.

Our measurements of the valley splitting in a quantum point contact appear to be consistent
with the few-electron, few-step limit. The same conclusions should therefore also apply
to a small quantum dot geometry. We expect a quantum dot fabricated in an equivalent
quantum well to produce a valley splitting in the range 0.35-0.46~meV, depending on the size
of the dot. Additional control of the valley splitting can be achieved by varying the quantum
well width or the doping density. The former has a very strong effect on the valley splitting,
but can also lead to valley splitting oscillations \cite{boykin04}.
The latter essentially determines the
electric field on the quantum dot. We believe the doping density forms a more robust design
parameter, and may lead to valley splittings on the order of 1~meV. Additional details and
discussion are provided in the Supplementary Materials.

Finally, we probe the limit of weak lateral confinement, using a Hall bar geometry.
As shown in the inset of Fig.~5, Shubnikov de Haas oscillations in a 
two-dimensional electron gas reveal the lifting of both the spin and valley degeneracies 
at high magnetic field.  Previous results indicate that the valley splitting is smaller 
than the spin splitting in large structures like this Hall bar
\cite{weitz96,koester97,khrapai03,lai04,pudalov}, 
a result that we confirm here with microwave spectroscopy.

Microwave driven electron spin resonance can be detected by measuring the change in resistance of a Hall bar upon application of a microwave field \cite{jiang01}, so-called electrically-detected electron spin resonance (ED-ESR).  We have performed ED-ESR and observed the classic Zeeman spin splitting with a $g$-factor of 2, as expected for silicon.  Here we extend this technique and use microwave spectroscopy to measure the valley splitting.  The valley splitting values $E_v$ are smaller than the spin splitting, and are presented in Fig.~5 as a function of the
applied magnetic field.  The smaller magnitude of the valley splitting explains the weaker minima in the Shubnikov de Haas oscillations associated with the valley splitting.  
The microwave spectroscopy shows that the valley splitting is linear all the way down to 
0.3~T, with a slope of $24.7\pm0.4~\mu\text{eV/T}$.  The sharpness of the resonance peaks allows very tight error bars, with 
a small zero-field intercept of $E_v(0)=1.5\pm0.6~\mu\text{eV}$.  The slope and intercept are in general agreement with previous reports in Si/SiGe quantum wells \cite{weitz96} and Si inversion layers 
\cite{khrapai03}, but the present work extends those measurements down to much lower fields.

Qualitatively, the magnetic field dependence of the 2DEG data is consistent with
theory discussed above.  In a magnetic field, the 
electron is confined over the magnetic length scale $l_B=\sqrt{\hbar /|eB|}$.  Numerical investigations of the valley splitting in a 
stepped quantum well yield trends similar to Fig.~5 with valley splitting increasing as 
$B$ increases.  To obtain correspondence with the data,
we must introduce specific, though
plausible, disorder models for the step profiles. 
The simulations then predict slopes very similar to our valley resonance experiments,
and a valley splitting that essentially vanishes at zero field.  

We have shown here that valley splitting can be controlled through both physical and magnetic confinement.  The small valley splitting observed in numerous
previous experiments arises because disorder and steps in the quantum well suppress the valley splitting by interference.  As a consequence, strong lateral confinement can 
reduce the interference and increase valley splitting substantially.  It is interesting 
to note that these results now point to other ways of increasing the valley splitting.  For example, a smaller step density can be obtained by 
growing on substrates with little average tilt (\textit{e.g.}, in certain Si/SiO$_2$
devices \cite{takashina06}).
However, the thick growths required to obtain relaxed SiGe on Si substrates will still lead to steps arising from dislocation formation.  An alternative that requires no dislocations is the growth of two-dimensional 
electron gases in Si/SiGe nanomembranes \cite{roberts06}.  More generally, these results show that even properties that at first glance appear to be intrinsic and unalterable --- in large part a band structure phenomenon --- can in fact be tuned and controlled with methods of nanoscale fabrication. 

\section*{METHODS}
\section*{Materials Growth}
The Si/SiGe heterostructures used in these experiments were grown by
ultrahigh vacuum chemical vapor deposition \cite{ismail95}.  For each sample, the two-dimensional electron gas (2DEG)
is located in a strained Si quantum well grown on a
strain-relaxed Si$_{1-x}$Ge$_{x}$ buffer layer.  For the quantum point contact experiment, $x=0.25$, the quantum well width is $100 \textrm{\AA}$, and the
density and mobility are $n=5.7\times 10^{11}\, \text{cm}^{-2}$ and
$\mu=200,000\, \text{cm}^{2}$/Vs, respectively. 
For the sample used in the microwave measurements, $x=0.3$, the quantum well width is $80 \textrm{\AA}$, the 2DEG density
$n=4.2\times 10^{11}\, \text{cm}^{-2}$, and the mobility $\mu=40,000\, \text{cm}^{2}$/Vs. 
Further details about the structures can be found in
Ref.~\onlinecite{klein04}.

\section*{SUPPLEMENTARY MATERIAL}

\section*{Fabrication}
A variety of methods have been used to fabricate nanostructures in silicon/silicon-germanium 
devices.  These include etching, Schottky top and side gates, and combinations of these 
approaches \cite{toebben95,wieser02,klein04,sakr05,keith05,berer06,scappucci06}.  Here, we 
use Schottky top-gates formed by the evaporation of palladium onto etched mesas.  Ohmic 
contact to the two-dimensional electron gas is achieved by annealing an evaporated Au/Sb film.

\section*{Conductance Fitting}
Here, we describe methods used to obtain the the valley splitting shown in Fig.~3 of
the main text.  Similar results are shown in Fig.~S1, here, for the second subband of the quantum point contact.  The procedure we use follows the canonical method of 
Ref.~\onlinecite{vanwees88}.  We map the conductance at for all magnetic fields and 
gate voltages to integer values of the conductance quantum $e^2/h$.  Transitions to the 
next conductance plateau occur when a subband energy given by Eq.~4 is equal to the Fermi energy.   The values $n=1,2 \text{, } n_B=\pm1/2 \text{, and } n_v=\pm1/2$ determine the 
Landau level for a given subband energy, reflecting the orbital level, and the spin and 
valley states, respectively.  We set $\hbar\omega_0=0$, leaving only five fitting parameters, 
$\Delta_\text{ext}$ and $\Delta_B$ (for $n=1,2$) and $V_b$.  Because there are more 
constraint equations than fitting parameters, for any given gate voltage, we use a 
least squares method to determine the optimum fit.

\section*{Microwave Spectroscopy}
A schematic of the Hall bar used for electron valley resonance measurements is shown in 
Fig.~S2.  A double lock-in technique
is used to measure the change in resistance $\Delta R_{xx}$ of the 2DEG 
as a function of the perpendicular field
$B$, in the presence of microwaves. Lock-in 1
provides a bias current ranging from 100~nA to 250~nA, modulated at
701.3~Hz. Lock-in 2 is used to modulate the microwave amplitude
with 100\% modulation at 5.7~Hz. The output of Lock-in 1 is fed into
Lock-in 2, which measures $\Delta R_{xx}$.
Microwaves are produced by an HP83650A synthesizer, and are carried
down to the sample using a low loss coaxial line terminating about 5
cm from the surface of the sample in a loop antenna. The base of a
resonant cavity is replaced with a sample stage. 
The microwave power at the sample has a strong frequency
dependence because of the open cavity and the impedance mismatches along
the length of the coaxial line.  Because of this non-uniformity, a wide
range of powers (10 $\mu$W-10 mW) are used to ensure optimal
power delivery. The magnetic field is produced by a
superconducting magnet and all measurements are carried out in an
Oxford Instruments $^3$He cryostat with a base
temperature of 0.25~K.

\begin{figure}[t] 
  \centering
  \includegraphics[bb=57 295 497 576,width=3.3in,keepaspectratio]{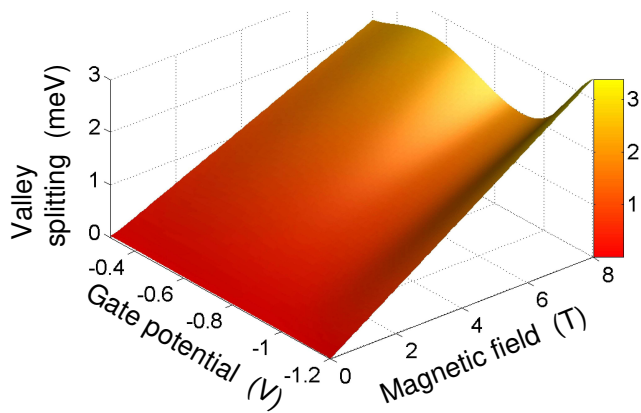}
  \caption{
The valley splitting in the second quantum level as a function of magnetic field $B$ and 
gate voltage $V_g$.}
  \label{fig:supfig1}
\end{figure}

The same experimental set-up can be used to detect both ESR and EVR
signals. We observe
typical ESR resonances, with linewidths on the order of 5~G. The
EVR transition is slightly different than ESR because it
is not driven by magnetic fields.  (The two low-lying valley states 
are orthogonal and unaffected by the spin operator, causing the 
Zeeman transition matrix element to vanish.) 
However, the electric dipole transition is allowed by general symmetry considerations 
\cite{kleiner}, which apply to the quantum well geometry.

To analyze the valley resonance features, we fit the data.  First, the background 
resistance is removed by fitting to a second degree polynomial away
from the main peak.  We find that Gaussians provide the best representation of the 
individual peaks, with peak widths on the order of 20-25~G.  Typically, the 
resonance features account for about one part in $10^4$  
of the total resistance signal.  
The fitted peak positions are plotted
in Fig.~2 of the main text, as a function of the perpendicular magnetic field. 

\section*{Valley Splitting in a Quantum Dot}
In Fig.~3 of the main text, large negative gate potentials correspond to the 
pinch-off regime, where the 2DEG below the gates is almost fully depleted.
In this regime, the electron density is low, and the channel is so narrow that the 
magnetic field provides no additional confinement.  
Consequently, the valley splitting is independent of magnetic field, with a constant
value of about $0.35$~meV.  

\begin{figure}[t] 
  \centering
  \includegraphics[bb=43 212 577 706,width=3.1in,keepaspectratio]{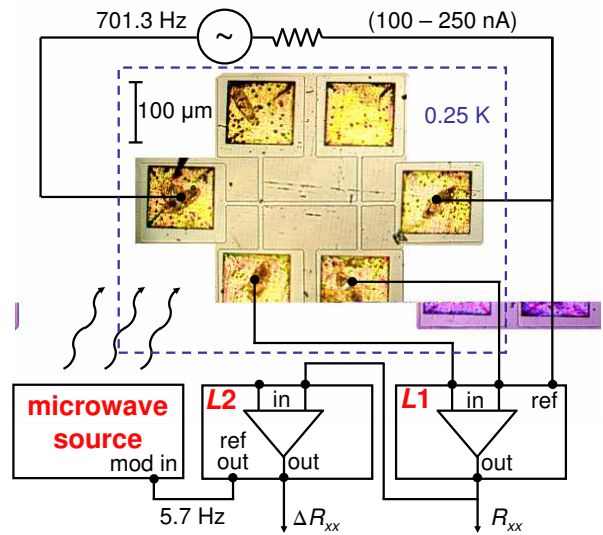}
  \caption{
Experimental schematic of the microwave valley resonance experiment showing the two lock-in amplifiers $L1$ and $L2$.}
  \label{fig:supfig2}
\end{figure}

We can compare this experimental value of the valley splitting with theoretical expectations,
based on the effective mass theory of valley coupling described in 
Ref.~\onlinecite{friesenunp}.  For a 
point contact device, we assume that the electric field in the quantum well
is largely determined by the doping density $n_d$ (justification is given
below).  The electric field from a fully ionized doping layer is given by
$E=en_d/2\varepsilon$.  We consider the limit of a narrow quantum point contact, which sees
no steps.  The theoretical expression for the valley splitting is then
\begin{equation}
E_v \simeq \left\{
\begin{array}{ll}
4v_v \left| e^{ik_0L}(x-y)+e^{-ik_0L}(x+y) \right|/L & \quad (x>y) \\
8v_vy/L & \quad (x\le y) ,
\end{array} \right. \label{eq:EvwithE}
\end{equation}
where
\begin{equation}
x = \frac{\hbar^2\pi^2}{2m_lL^2\Delta E_c} \quad \text{and} \quad
y = \frac{eEL}{4\Delta E_c}. \label{eq:VS}
\end{equation}
Here, $m_l=0.91m_0$ is the longitudinal electron effective mass in silicon, 
$k_0=9.5\times 10^9$~m$^{-1}$ is the silicon valley wavevector, $L$ is the thickness
of the quantum well, $\Delta E_c$ is the conduction band offset for the 
silicon/silicon-germanium
quantum well, and $v_v=\Delta E_c \times 5.0\times 10^{-11}$ is the valley coupling 
constant, obtained from a many-band tight binding calculation \cite{friesenunp},
such as Ref.~\onlinecite{boykin04}.  (Here, $v_v$ is in units of 
eV$\cdot$m when $\Delta E_c$ is in units of eV.)  The  
device parameters that determine the valley splitting are therefore the band offset, the
thickness of the quantum well, and the doping density. 

Figure~S3 shows the valley splitting obtained from Eq.~(\ref{eq:VS}),
for a heterostructure with 30\% germanium barriers.  The device parameters
consistent with our quantum point contact sample are indicated 
as a white dot, giving $E_v=0.46$~meV as the theoretical prediction for the
valley splitting.  Comparing this value to the
experimentally measured valley splitting at pinch-off gives very reasonable
agreement.  The experimental value is slightly suppressed below the  
theoretical prediction, due to the presence of a small number of interfacial steps in 
the active region of the device.  Thus, we conclude that the pinch-off regime
of the quantum point contact corresponds to the few-step limit, in
which valley splitting approaches its theoretical upper bound.  The same limit is
appropriate for small quantum dots, of interest for quantum computing. 
Although the quantum dot geometry is slightly different than the quantum point contact, 
the experimental parameters that determine the valley splitting are the same. 

We are now in the position to discuss device parameters that enable large
valley splittings in quantum dots.  As explained above, Fig.~S3 provides a good 
starting point for predicting the behavior of small, single electron dots.  
We see that a large valley splitting
can be achieved either by decreasing the quantum well width, or by increasing the doping
density.  In the first case, we notice oscillations in the valley splitting,
in the lower left-hand portion of the figure.  These arise from interference effects,
when the electron wavefunction has significant amplitude on both the top and 
bottom interfaces of the quantum well \cite{friesenunp,boykin04}.  While it is 
possible to achieve very large valley splittings in this regime, we note that
fluctuations of a fraction of a nanometer in the growth thickness can result in a 
strong suppression of the valley splitting, due to the oscillations.  This narrow
quantum well regime also has reduced mobility, and is therefore undesirable for
building devices.

In the second case, increasing the dopant charge density will increase the electric field at the quantum well if all the dopants are ionized and fall into the well.  This is ensured by decreasing the setback distance between the dopant layer and the quantum well as the dopant density is increased.  It will be important that the dopant layer is not too distant from the dot, so that edge effects do not overly reduce the electric field in the direction normal to the quantum well.
As long as the electric field is large enough, and the quantum well is not too narrow, the valley splitting will be large and without oscillations, corresponding to the 
region above the dashed line in Fig.~S3.  

\begin{figure}[t] 
  \centering
  \includegraphics[bb=45 463 425 761,width=3in,keepaspectratio]{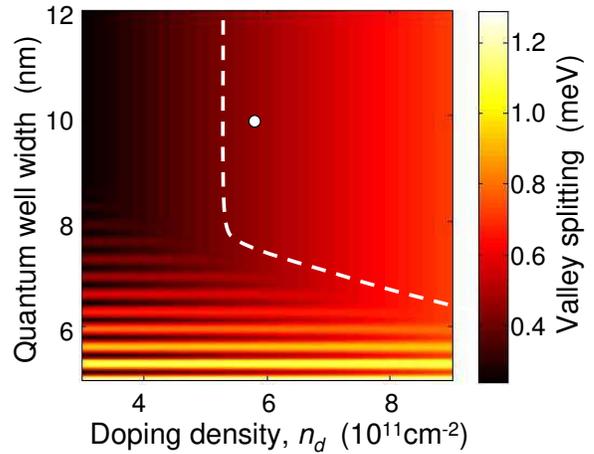}
  \caption{
Valley splitting calculation for a quantum dot, as a function of doping 
density and quantum well thickness.  The white dot corresponds to the heterostructure
used for the quantum point contact, giving a valley splitting of 0.46~meV.  
The dashed line marks the target region for quantum dot samples.}
  \label{fig:supfig3}
\end{figure}

\section*{ACKNOWLEDGMENTS}
We gratefully acknowledge conversations with R. Blick.
This work was supported by NSA/LPS under ARO contract  
number W911NF-04-1-0389, and by the National Science Foundation through the ITR 
program (DMR-0325634) and the EMT program (CCF-0523675).

\end{document}